\documentclass[12pt,epsf,fleqn]{article}
\usepackage{epsfig}
\usepackage[dvips]{color}
\begin{document}
\title{Explicit CP violation in a MSSM with an extra $U(1)'$}
\author{S.W. Ham$^{(1)}$, E.J. Yoo$^{(2)}$, and S.K. Oh$^{(1,2)}$
\\
\\
{\it $^{(1)}$ Center for High Energy Physics, Kyungpook National University}
\\
{\it Daegu 702-701, Korea}
\\
{\it $^{(2)}$ Department of Physics, Konkuk University, Seoul 143-701,
Korea}
\\
\\
}
\date{}
\maketitle
\begin{abstract}
We study that a minimal supersymmetric standard model with an extra $U(1)'$ gauge symmetry may
accommodate the explicit CP violation at the one-loop level through radiative corrections.
This model is CP conserving at the tree level and cannot realize the spontaneous CP violation
for a wide parameter space at the one-loop level.
In explicit CP violation scenario, we calculate the Higgs boson masses and the magnitude of
the scalar-pseudoscalar mixings in this model at the one-loop level
by taking into account the contributions of top quarks, bottom quarks, exotic quarks,
and their superpartners.
In particular, we investigate how the exotic quarks and squarks would affect the scalar-pseudoscalar mixings.
It is observed that the size of the mixing between the heaviest scalar and pseudoscalar Higgs bosons
is changed up to 20 \% by a complex phase originated from the exotic quark sector of this model.
\end{abstract}
\vfil\eject

\section{INTRODUCTION}

Possessing two Higgs doublets is one of the characteristics of
the minimal supersymmetric standard model (MSSM)
in order to give masses to the up-quark sector and the down-quark sector separately [1].
This property enables in principle CP violation to occur
in the Higgs sector of the MSSM through the mixing between the CP even and the CP odd states [2].
The mechanism of CP violation in the MSSM has been studied by many authors [3-12].
It is found that at the tree level, neither explicit nor spontaneous CP violation is possible
in the Higgs sector of the MSSM, because any complex phases in it can always be eliminated
by rotating the Higgs fields.
Even at the one-loop level, spontaneous CP violation is disfavored
because it requires a very light neutral Higgs boson, which has already been ruled out
by experiments [3,4].
On the other hand, explicit CP violation is viable in the MSSM at the one-loop level
by virtue of the radiative corrections due to the loops of relevant particles,
such as quarks and squarks [5-12].
The radiative corrections by these particles yield the mixing
between the CP even and the CP odd neutral Higgs bosons.
Thus, it is possible to achieve the explicit CP violation
in the radiatively corrected Higgs sector of the MSSM.

However, a drawback of the MSSM is that it contains a Higgs mixing term $\mu H_1 H_2$,
where $\mu$ is a parameter having mass dimension, and $H_i$ ($i=1,2$) are Higgs doublets.
It is known that this $\mu$ parameter causes a hierarchy problem
with respect to the symmetry breaking scale [13].
A natural framework to solve this problem is to generate the $\mu$ parameter
via the vacuum expectation value (VEV) of a Higgs singlet [14].
A possibility for the framework is to extend simply the Higgs sector of the MSSM
by introducing a Higgs singlet, like the next-to-minimal supersymmetric standard model,
where the $\mu$ parameter is practically replaced
by a dimensionless parameter times the VEV of the Higgs singlet [14].

A more plausible possibility to solve the $\mu$ problem is to extend not only the Higgs sector
but also the gauge symmetry of the MSSM by introducing a nonanomalous $U(1)'$ symmetry broken
at the TeV scale [15].
Extending the gauge symmetry to include an additional $U(1)'$ symmetry is widely considered
in various theoretical models, such as string models or GUT models.
The MSSM with an extra $U(1)'$ symmetry can actually forbid the $\mu H_1 H_2$ term
but allow $\lambda S H_1 H_2$, by assigning appropriate $U(1)'$ charges,
where $S$ is the Higgs singlet and $\lambda$ is a dimensionless parameter.
After $S$ develops a VEV of the electroweak scale, this $\lambda$ term would
effectively generate the $\mu$ term at the electroweak scale [16].

In the literature, a number of studies have been performed on the MSSM with an extra $U(1)'$ symmetry [17-20].
Several authors have calculated the Higgs boson masses from the tree-level Higgs potential of this model [17]
as well as from the radiatively corrected Higgs potential [18,19].
It is found that, at the tree level, the lightest scalar Higgs boson in this model
can be heavier than $Z$ boson.
That is, the upper bound on the mass of the lightest scalar Higgs boson in this model
is larger than the one in the MSSM without an extra $U(1)'$ symmetry.
Moreover, it is also found that this tree-level mass of the lightest scalar Higgs boson in this model
can be significantly affected by radiative corrections.

Recently, this model has been further investigated within the context of the supersymmetric CP problem [20],
as it can have explicit CP violating phases in the Higgs sector.
At the tree level, like the MSSM, this model has no CP mixing between the CP even and the CP odd states
in the Higgs sector because the single CP phase arising from the tree-level Higgs potential
can always be eliminated by rotating the Higgs fields.
In Ref. [20], the one-loop contribution from the top and stop quark loops are considered
in the explicit CP violation scenario.
It is observed that the lightest neutral Higgs boson remains essentially CP even for a wide parameter space,
with a fixed value of $\tan \beta = 1$, where $\tan \beta = v_2/v_1$ is the ratio of the VEVs of two Higgs doublets.
Meanwhile, large CP mixtures for the other heavier Higgs bosons may be realized as the size of the effective
$\mu$ terms becomes large.

In this paper, motivated by the results of Ref. [20], we would like to study this model
in some detail in the explicit CP violation scenario.
The radiative CP mixing may be generated by means of the complex phases coming from
the one-loop effective potential due to top quarks, bottom quarks, the exotic quarks,
and their superparticles, where the exotic quarks and squarks with electric charges $\pm 1/3$
are introduced into this model in order to cancel gauge anomaly [21].
In particular, we are interested in the contributions of exotic quark sector in this model.
At the one-loop level, the exotic quarks and squarks are found to play some recognizable role
in the scalar-pseudoscalar mixings.
We find that the relevant CP phase from the exotic quark sector may change the size of the CP mixing
between the heaviest scalar and the pseudoscalar Higgs bosons by up to 20 \%.
Also we find that this model yields a negative or extremely small axion mass,
as well as other unacceptable predictions, at the one-loop level for full parameter space
in the spontaneous CP violation scenario.
Thus, we note that the present model is impossible to realize spontaneous CP violation at the one-loop level.

\section{HIGGS SECTOR}

GUT, such as the string-inspired $E_6$ model, might be a natural motivation
to enlarge the gauge symmetry in order to accommodate an extra $U(1)'$ symmetry into the MSSM.
The $E_6$ gauge group may be decomposed into
\begin{equation}
    E_6 \supset SU(10)\times U(1)_{\psi} \supset SU(5) \times U(1)_{\chi} \times U(1)_{\psi} \ ,
\end{equation}
where $SU(5)$ is further broken down to the Standard Model (SM) gauge group,
$SU(3)_C \times SU(2)_L \times U(1)_Y$.
At the electroweak scale, the desired extra $U(1)'$ symmetry may be given
as an orthogonal linear combination of $U(1)_{\chi}$ and $U(1)_{\psi}$ as
\begin{equation}
    Q' = \cos \theta_E Q_{\chi} + \sin \theta_E Q_{\psi} \ ,
\end{equation}
where $Q'$, $Q_{\chi}$, and $Q_{\psi}$ are the $U(1)'$, $U(1)_{\chi}$,
and $U(1)_{\psi}$ charges, respectively.
Sometimes, for four different values of the angle $\theta_E$,
the four particular combinations of the $U(1)_{\chi}$ and $U(1)_{\psi}$ are called as follows:
the $\chi$-model for $\theta_E = 0$, the $\psi$-model for $\theta_E = \pi/2$,
the $\eta$-model for $\theta_E = \tan^{-1} (-\sqrt{5/3})$,
and the $\nu$-model for $\theta_E = \tan^{-1} \sqrt{15}$.
The gauge group we consider is thus $G = SU(3)_C \times SU(2)_L \times U(1)_Y \times U(1)'$.

The Higgs sector of the MSSM with an extra $U(1)'$ symmetry consists of
two Higgs doublets $H_1 = (H_1^0, H_1^-)$ and $H_2 = (H_2^+, H_2^0)$, and a neutral Higgs singlet $S$.
This model possesses an extra pair of $SU(2)$ singlet quarks, $D_L$ and ${\bar D}_R$,
with electric charges $-1/3$ and $+1/3$, respectively, introduced
by the requirement of the gauge anomaly cancellation.
For the fermion matter fields in this model,
we take into account only the third generation of quarks, besides the exotic quarks.
Then, the superpotential for the model we consider may be expressed as
\begin{equation}
    W \approx \lambda S H_1 H_2 + h_t Q H_2 t_R^c + h_b Q H_1 b_R^c + k S D_L {\bar D}_R   ,
\end{equation}
where $Q$ is the left-handed quark doublet, $t_R^c$ and $b_R^c$ are the charge conjugate
of the right-handed top quark and bottom quark, respectively, with $h_t$ and $h_b$ being
their respective Yukawa coupling coefficients, and $H_1 H_2 = H_1^0 H_2^0 - H_1^- H_2^+$.
All the coupling coefficients are dimensionless.
This superpotential is symmetric under $SU(3)_C \times SU(2)_L \times U(1)_Y \times U(1)'$,
and the $U(1)'$ symmetry is broken down by the VEV of $S$.

The Higgs potential of this model at the tree level may expressed as a sum of the $F$-term,
the $D$-term, and the soft breaking term, that is,
\begin{equation}
    V_0 = V_F + V_D + V_{\rm S} \ ,
\end{equation}
where
\begin{eqnarray}
V_F & = & |\lambda|^2 [(|H_1|^2 + |H_2|^2) |S|^2 + |H_1 H_2|^2]  \ , \cr
V_D & = & {g_2^2 \over 8} (H_1^{\dagger} \vec\sigma H_1 + H_2^{\dagger} \vec\sigma H_2)^2
+ {g_1^2 \over 8} (|H_1|^2 - |H_2|^2)^2 \cr
& &\mbox{}+ {g^{'2}_1 \over 2} ( {\tilde Q}_1 |H_1|^2 + {\tilde Q}_2 |H_2|^2 + {\tilde Q}_3 |S|^2)^2 \ , \cr
V_{\rm S} & = & m_1^2 |H_1|^2 + m_2^2 |H_2|^2 + m_3^2 |S|^2
- [\lambda A_{\lambda} (H_1 H_2) S + {\rm H.c.}] \ ,
\end{eqnarray}
where $m_i^2$ ($i$ = 1,2,3) are soft masses,  $g_2$, $g_1$, and $g_1'$ are respectively
the coupling coefficients of $SU(2)$, $U(1)$, and $U(1)'$; $\vec \sigma = (\sigma_1, \sigma_2, \sigma_3)$
are the Pauli matrices; ${\tilde Q}_1$, ${\tilde Q}_2$, and ${\tilde Q}_3$ are respectively
the $U(1)'$ charges of $H_1$, $H_2$, and $S$,
satisfying the condition of ${\tilde Q}_1 + {\tilde Q}_2 + {\tilde Q}_3 = 0$
to obey the gauge invariance of the superpotential under $U(1)'$.

At the electroweak scale, the three neutral Higgs fields develop VEVs
as $<H_1^0> = v_1$, $<H_2^0> = v_2$, and $<S> = s e^{i \phi_s}$,
where $\phi_s$ is the relative phase between $H_1 H_2$ and $S$.
In general, $\lambda A_{\lambda}$ in the soft breaking term may be complex,
like $\lambda A_{\lambda} e^{\phi}$.
However, it is always possible to make the phase $\phi$ and
the phase $\phi_s$ cancel out each other by redefining the Higgs singlet $S$,
and thus the tree-level Higgs potential can be made free of any complex phase.
Therefore, the CP symmetry is not violated in this model at the tree level.

Now let us consider the one-loop radiative corrections to the tree-level Higgs potential.
In supersymmetric models, the incomplete cancellation between ordinary particles
and their superpartners yield the one-loop corrections to the tree-level Higgs boson masses.
Generally, the most dominant part of the one-loop corrections to the tree-level Higgs potential
come primarily from the top and stop quark loops.
By considering the top quark sector only, it has been observed that explicit CP violation
in this model is viable through the radiatively corrected Higgs potential [20].
For large $\tan \beta$, the contribution of the bottom and sbottom quark loops
can also be significantly large.
We would like to consider in this paper the one-loop contributions from both the top and bottom quark sector,
as well as from the exotic quark sector.

To start with, we calculate the tree-level masses of the relevant particles
in order to evaluate the radiative corrections
to the neutral Higgs sector.
The fermion matter fields obtain their masses after the electroweak symmetry breaking
as $m_t^2 = (h_t v_2)^2$ for top quark, $m_b^2 = (h_b v_1)^2$ for bottom quark, and $m_k^2 = (k s)^2$ for the exotic quark.
The tree-level masses of their superpartners are given as
\begin{eqnarray}
m_{{\tilde t}_1, \ {\tilde t}_2}^2 & = & m_T^2 + m_t^2 \mp m_t
\sqrt{A_t^2 + \lambda^2 s^2 \cot^2 \beta - 2 \lambda A_t s \cot \beta \cos \phi_t} \ , \cr
m_{{\tilde b}_1, \ {\tilde b}_2}^2 & = & m_B^2 + m_b^2 \mp m_b
\sqrt{A_b^2 + \lambda^2 s^2 \tan^2 \beta - 2 \lambda A_b s \tan \beta \cos \phi_b} \ , \cr
m_{{\tilde k}_1, \ {\tilde k}_2}^2 & = & m_K^2 + m_k^2 \mp m_k
\sqrt{A_k^2 + \lambda^2 v^4 \sin^2 2 \beta/(4 s^2) - \lambda A_k v^2 \sin 2 \beta \cos \phi_k/s} \ ,
\end{eqnarray}
where $\tan\beta = v_2/v_1$, $v^2 = v^2_1 + v^2_2$.
Further, $m_T$, $m_B$, and $m_K$ are the soft SUSY breaking masses respectively for stop quarks,
sbottom quarks, and the exotic squarks, and likewise $A_t$, $A_b$, and $A_k$ are respectively
their trilinear soft SUSY breaking parameters of the mass dimension.
One can notice that there are three phases $\phi_t$, $\phi_b$, and $\phi_k$ in the above expressions.
These complex phases are determined by the generally complex $A_q$ ($q = t, b, k$)
and the phase of $s$, the VEV of the Higgs singlet.
We note that the $D$-terms do not contribute to the squark masses.

The full Higgs potential at the one-loop level may be written as
\[
    V = V_0 + V_1 \ ,
\]
where $V_1$ is the radiative corrections due to the relevant particles and their superpartners.
According to the effective potential method [22], $V_1$ is given as
\begin{equation}
V_1  = \sum_{l} {n_l {\cal M}_l^4 \over 64 \pi^2}
\left [
\log {{\cal M}_l^2 \over \Lambda^2} - {3 \over 2}
\right ]  \ ,
\end{equation}
where $\Lambda$ is the renormalization scale in the modified minimal subtraction scheme,
and the subscript $l$ stands for various participating particles: $t$, $b$, $k$,
${\tilde t}_1$, ${\tilde t}_2$, ${\tilde b}_1$, ${\tilde b}_2$, ${\tilde k}_1$ and ${\tilde k}_2$.
The degrees of freedom for these particles including the sign convention are $n_t = n_b = n_k = -12$ and
$n_{{\tilde t}_i} = n_{{\tilde b}_i} = n_{{\tilde k}_i} = 6$ ($i=1,2$),
since in the above formula enter fermions with a negative sign while bosons with a positive sign.

If CP violation takes place in the Higgs sector, the neutral Higgs bosons
would not have definite CP parity, hence mixings among them.
In case of explicit CP violation, the non-trivial tadpole minimum condition
with respect to the pseudoscalar component of the Higgs field is given as
\begin{eqnarray}
0 & = & A_{\lambda} \sin \phi_0
- {3 m_t^2 A_t \sin \phi_t \over 16 \pi^2 v^2 \sin^2 \beta} f (m_{{\tilde t}_1}^2,  \ m_{{\tilde t}_2}^2)
- {3 m_b^2 A_b \sin \phi_b \over 16 \pi^2 v^2 \cos^2 \beta} f (m_{{\tilde b}_1}^2,  \ m_{{\tilde b}_2}^2) \cr
&&\mbox{} - {3 m_k^2 A_k \sin \phi_k \over 16 \pi^2 s^2} f (m_{{\tilde k}_1}^2,  \ m_{{\tilde k}_2}^2)  \ ,
\end{eqnarray}
where the dimensionless function $f$ arising from radiative corrections is defined as
\[
 f(m_x^2, \ m_y^2) = {1 \over (m_y^2 - m_x^2)} \left[  m_x^2 \log {m_x^2 \over \Lambda^2} - m_y^2
\log {m_y^2 \over \Lambda^2} \right] + 1 \ ,
\]
and the phase $\phi_0$ is given by $\phi_0 = \phi_s + \phi$, with $\phi$ being the phase of
$\lambda A_{\lambda}$ in $V_0$ and $\phi_s$ being the phase of $s$.
Note that $\phi_0$ is generally nonzero at the one-loop level
whereas at the tree level it can be made zero by rotating the Higgs singlet.
In the above tadpole minimum condition, the first term comes from the tree-level Higgs potential,
and the remaining three terms come respectively from the top squark, bottom squark, and exotic squark contributions.

At the tree level, the Higgs sector in the MSSM with an extra $U(1)'$ has ten real degrees of freedom,
which are decomposed by two neutral Goldstone bosons, a pair of charged Goldstone bosons,
four neutral Higgs bosons and a pair of charged Higgs bosons.
After the electroweak symmetry breaking, the two neutral Goldstone bosons and
a pair of charged Goldstone bosons will eventually be absorbed
into the longitudinal component of $Z$, $Z'$, and $W$ gauge bosons.
Since the CP symmetry is conserved in the Higgs sector at the tree level,
the four neutral Higgs bosons can be divided into three scalar Higgs bosons
and one pseudoscalar Higgs boson, according to the CP parity.

The squared mass matrix $M$ of the four neutral Higgs bosons is given
as a symmetric $4 \times 4$ matrix that is obtained by the second derivatives of the Higgs potential
with respect to the four Higgs fields, in the ($h_1, h_2, h_3, h_4$) basis.
At the tree level, $M$ is given by $V^0$ as
\begin{equation}
     M = M^0 =
    \left ( \begin{array}{cccc}
    M_{11}^0 & M_{12}^0 & M_{13}^0 & 0  \cr
    M_{21}^0 & M_{22}^0 & M_{23}^0 & 0  \cr
    M_{31}^0 & M_{32}^0 & M_{33}^0 & 0  \cr
    0 & 0 & 0 & m_A^2
        \end{array}
    \right ) \ ,
\end{equation}
where the tree-level squared mass of the pseudoscalar Higgs boson is given as
\[
    m_A^2  = 2 \lambda A_{\lambda} v {\cos \phi_0  \over \sin 2 \alpha}
\]
with
\[
    \tan \alpha = {v \over 2 s} \sin 2 \beta
\]
standing for the splitting between an extra $U(1)'$ symmetry breaking scale and the electroweak scale.
Note that $M^0$ is divided into two blocks of the upper-left $3 \times 3$ submatrix and
the single element $M_{44}^0 = m_A^2$, corresponding to the scalar part and the pseudoscalar part.
There is no scalar-psuedoscalar mixing in $M^0$, since $M^0_{i4}$ ($i=1-3$) are zero, and hence no CP violation.
Explicit calculations yield
\begin{eqnarray}
M_{11}^0 & = & m_Z^2 \cos^2 \beta + 2 g^{'2}_1 {\tilde Q}_1^2 v^2 \cos^2 \beta
+ m_A^2 \sin^2 \beta \cos^2 \alpha  \ ,  \cr
M_{22}^0 & = & m_Z^2 \sin^2 \beta + 2 g^{'2}_1 {\tilde Q}_2^2 v^2 \sin^2 \beta
+ m_A^2 \cos^2 \beta \cos^2 \alpha \ ,  \cr
M_{33}^0 & = & 2 g^{'2}_1 {\tilde Q}_3^2 s^2 +A^2 \sin^2 \alpha \ , \cr
M_{12}^0 & = & g^{'2}_1 {\tilde Q}_1 {\tilde Q}_2 v^2 \sin 2 \beta + (\lambda^2 v^2 - m_Z^2/2) \sin 2 \beta
- m_A^2 \cos \beta \sin \beta \cos^2 \alpha \ ,  \cr
M_{13}^0 & = & 2 g^{'2}_1 {\tilde Q}_1 {\tilde Q}_3 v s \cos \beta + 2 \lambda^2 v s \cos \beta
- m_A^2 \sin \beta \cos \alpha \sin \alpha \ , \cr
M_{23}^0 & = & 2 g^{'2}_1 {\tilde Q}_2 {\tilde Q}_3 v s \sin \beta + 2 \lambda^2 v s \sin \beta
- m_A^2 \cos \beta \cos \alpha \sin \alpha  \ ,
\end{eqnarray}

At the one-loop level, the squared mass matrix may be decomposed as
\[
    M = {\bar M}^0 + M^1
\]
where ${\bar M}^0$ has exactly the same appearance as $M^0$ but $m_A^2$ in $M^0$ should be replaced
by ${\bar m}_A^2$ in ${\bar M}^0$, where ${\bar m}_A^2$ is given as
\begin{eqnarray}
{\bar m}_A^2 & = & {2 \lambda v \over \sin 2 \alpha} \left [A_{\lambda} \cos \phi_0
- {3 m_t^2 A_t \cos \phi_t \over 16 \pi^2 v^2 \sin^2 \beta} f (m_{{\tilde t}_1}^2,  \ m_{{\tilde t}_2}^2)
\right. \cr
& &\mbox{} \left.
- {3 m_b^2 A_b \cos \phi_b \over 16 \pi^2 v^2 \cos^2 \beta} f (m_{{\tilde b}_1}^2,  \ m_{{\tilde b}_2}^2)
- {3 m_k^2 A_k \cos \phi_k \over 16 \pi^2 s^2} f (m_{{\tilde k}_1}^2,  \ m_{{\tilde k}_2}^2) \right ] \ ,
\end{eqnarray}
One may notice that in the expression for ${\bar m}_A^2$,
the first term comes from the tree-level Higgs potential
and the remanning three terms comes from the contributions of the squarks.
Therefore, ${\bar M}^0$ contains not only the tree-level results
but also the contributions from radiative corrections.

Now, let us calculate $M^1$, which is obtained from $V_1$.
We may conveniently decompose $M^1$ as
\[
    M^1 = M^t + M^b + M^k
\]
where symmetric matrices $M^t$, $M^b$, and $M^k$ are obtained respectively from the top quark sector,
the bottom quark sector, and the exotic quark sector contributions to $V_1$,
after imposing the tadpole minimum condition.

Somewhat lengthy calculations yield the matrix elements of $M^t$ as
\begin{eqnarray}
    M_{11}^t & = & {3 m_t^4 \lambda^2 s^2 \Delta_{{\tilde t}_1}^2 \over 8 \pi^2  v^2 \sin^2 \beta}
{g(m_{{\tilde t}_1}^2, \ m_{{\tilde t}_2}^2) \over (m_{{\tilde t}_2}^2 - m_{{\tilde t}_1}^2)^2}  \ , \cr
    M_{22}^t & = & {3 m_t^4 A_t^2 \Delta_{{\tilde t}_2}^2 \over 8 \pi^2  v^2 \sin^2 \beta}
{g(m_{{\tilde t}_1}^2, \ m_{{\tilde t}_2}^2) \over (m_{{\tilde t}_2}^2 - m_{{\tilde t}_1}^2)^2}
 + {3 m_t^4 A_t \Delta_{{\tilde t}_2} \over 4 \pi^2 v^2 \sin^2 \beta}
{\log (m_{{\tilde t}_2}^2 / m_{{\tilde t}_1}^2) \over (m_{{\tilde t}_2}^2 - m_{\tilde{t}_1}^2)} \cr
& &\mbox{} + {3 m_t^4 \over 8 \pi^2 v^2 \sin^2 \beta}
\log \left ( {m_{{\tilde t}_1}^2  m_{{\tilde t}_2}^2 \over m_t^4} \right ) \ , \cr
    M_{33}^t & = & {3 m_t^4 \lambda^2 \Delta_{{\tilde t}_1}^2 \over 8 \pi^2 \tan^2 \beta}
{g(m_{{\tilde t}_1}^2, \ m_{{\tilde t}_2}^2) \over (m_{{\tilde t}_2}^2 - m_{{\tilde t}_1}^2 )^2} \ , \cr
    M_{44}^t & = & {3 m_t^4 \lambda^2 A_t^2 s^2 \sin^2 \phi_t \over 8 \pi^2 v^2 \sin^4 \beta \cos^2 \alpha}
{g(m_{{\tilde t}_1}^2, \ m_{{\tilde t}_2}^2) \over (m_{{\tilde t}_2}^2 - m_{{\tilde t}_1}^2 )^2}  \ , \cr
    M_{12}^t & = &\mbox{} - {3 m_t^4 \lambda A_t s \Delta_{{\tilde t}_1} \Delta_{{\tilde t}_2}
\over 8 \pi^2 v^2 \sin^2 \beta}
{g(m_{{\tilde t}_1}^2, \ m_{{\tilde t}_2}^2) \over (m_{{\tilde t}_2}^2 - m_{{\tilde t}_1}^2)^2}
- {3 m_t^4 \lambda s \Delta_{{\tilde t}_1} \over 8 \pi^2 v^2 \sin^2 \beta}
{\log (m_{{\tilde t}_2}^2 / m_{{\tilde t}_1}^2) \over (m_{{\tilde t}_2}^2 - m_{{\tilde t}_1}^2)} \cr
    M_{13}^t & = & {3 m_t^4 \lambda^2 s \Delta_{{\tilde t}_1}^2 \over 8 \pi^2 v \sin \beta \tan \beta}
{g(m_{{\tilde t}_1}^2, \ m_{{\tilde t}_2}^2) \over (m_{{\tilde t}_2}^2 - m_{{\tilde t}_1}^2)^2 }
- {3 m_t^2 \lambda^2 s \cos \beta \over 8 \pi^2 v \sin^2 \beta}
f(m_{{\tilde t}_1}^2, \ m_{{\tilde t}_2}^2) , \cr
    M_{14}^t & = & \mbox{} - {3 m_t^4 \lambda^2 A_t s^2 \Delta_{{\tilde t}_1} \sin \phi_t
\over 8 \pi^2 v^2 \sin^3 \beta \cos \alpha}
{g(m_{{\tilde t}_1}^2, \ m_{{\tilde t}_2}^2) \over (m_{{\tilde t}_2}^2 - m_{{\tilde t}_1}^2)^2 } \ , \cr
    M_{23}^t & = &\mbox{} - {3 m_t^4 \lambda A_t \Delta_{{\tilde t}_1} \Delta_{{\tilde t}_2}
\over 8 \pi^2 v \sin \beta \tan \beta}
{g(m_{{\tilde t}_1}^2, \ m_{{\tilde t}_2}^2) \over (m_{{\tilde t}_2}^2 - m_{{\tilde t}_1}^2)^2}
- {3 m_t^4 \lambda \cos \beta \Delta_{{\tilde t}_1} \over 8 \pi^2 v \sin^2 \beta}
{\log (m_{{\tilde t}_2}^2 / m_{{\tilde t}_1}^2) \over (m_{{\tilde t}_2}^2 - m_{{\tilde t}_1}^2) }  , \cr
    M_{24}^t & = & \mbox{} {3 m_t^4 \lambda A_t^2 s \Delta_{{\tilde t}_2} \sin \phi_t
\over 8 \pi^2 v^2 \sin^3 \beta \cos \alpha}
{g(m_{{\tilde t}_1}^2, \ m_{{\tilde t}_2}^2) \over (m_{{\tilde t}_2}^2 - m_{{\tilde t}_1}^2)^2 }
+ {3 m_t^4 \lambda A_t s \sin \phi_t \over 8 \pi^2 v^2 \sin^3 \beta \cos \alpha}
{\log (m_{{\tilde t}_2}^2 / m_{{\tilde t}_1}^2) \over (m_{{\tilde t}_2}^2 - m_{{\tilde t}_1}^2)}  \ , \cr
    M_{34}^t & = & \mbox{} - {3 m_t^4 \lambda^2 A_t s \Delta_{{\tilde t}_1} \sin \phi_t
\over 8 \pi^2 v \sin^2 \beta \tan \beta \cos \alpha}
{g(m_{{\tilde t}_1}^2, \ m_{{\tilde t}_2}^2) \over (m_{{\tilde t}_2}^2 - m_{{\tilde t}_1}^2)^2 } \ ,
\end{eqnarray}
where
\begin{eqnarray}
 \Delta_{{\tilde t}_1} & = & A_t \cos \phi_t - \lambda s \cot \beta  \  , \cr
 \Delta_{{\tilde t}_2} & = & A_t - \lambda s \cot \beta \cos \phi_t \ ,
\end{eqnarray}
and the dimensionless function $g$ is defined as
\[
 g(m_x^2,m_y^2) = {m_y^2 + m_x^2 \over m_x^2 - m_y^2} \log {m_y^2 \over m_x^2} + 2 \ .
\]

Likewise, the matrix elements of $M^b$ as
\begin{eqnarray}
    M_{11}^b & = & {3 m_b^4 A_b^2 \Delta_{{\tilde b}_1}^2 \over 8 \pi^2  v^2 \cos^2 \beta}
{g(m_{{\tilde b}_1}^2, \ m_{{\tilde b}_2}^2) \over (m_{{\tilde b}_2}^2 - m_{{\tilde b}_1}^2)^2}
 + {3 m_b^4 A_b \Delta_{{\tilde b}_1} \over 4 \pi^2 v^2 \cos^2 \beta}
{\log (m_{{\tilde b}_2}^2 / m_{{\tilde b}_1}^2) \over (m_{{\tilde b}_2}^2 - m_{{\tilde b}_1}^2)} \cr
& &\mbox{} + {3 m_b^4 \over 8 \pi^2 v^2 \cos^2 \beta}
\log \left ( {m_{{\tilde b}_1}^2  m_{{\tilde b}_2}^2 \over m_b^4} \right ) \ , \cr
    M_{22}^b & = & {3 m_b^4 \lambda^2 s^2 \Delta_{{\tilde b}_2}^2 \over 8 \pi^2  v^2 \cos^2 \beta}
{g(m_{{\tilde b}_1}^2, \ m_{{\tilde b}_2}^2) \over (m_{{\tilde b}_2}^2 - m_{{\tilde b}_1}^2)^2}  \ , \cr
    M_{33}^b & = & {3 m_b^4 \lambda^2 \Delta_{{\tilde b}_2}^2 \over 8 \pi^2 \cot^2 \beta}
{g(m_{{\tilde b}_1}^2, \ m_{{\tilde b}_2}^2) \over (m_{{\tilde b}_2}^2 - m_{{\tilde b}_1}^2 )^2} \ , \cr
    M_{44}^b & = & {3 m_b^4 \lambda^2 A_b^2 s^2 \sin^2 \phi_b \over 8 \pi^2 v^2 \cos^4 \beta \cos^2 \alpha}
{g(m_{{\tilde b}_1}^2, \ m_{{\tilde b}_2}^2) \over (m_{{\tilde b}_2}^2 - m_{{\tilde b}_1}^2 )^2}  \ , \cr
    M_{12}^b & = &\mbox{} - {3 m_b^4 \lambda A_b s \Delta_{{\tilde b}_1} \Delta_{{\tilde b}_2}
\over 8 \pi^2 v^2 \cos^2 \beta}
{g(m_{{\tilde b}_1}^2, \ m_{{\tilde b}_2}^2) \over (m_{{\tilde b}_2}^2 - m_{{\tilde b}_1}^2)^2}
- {3 m_b^4 \lambda s \Delta_{{\tilde b}_2} \over 8 \pi^2 v^2 \cos^2 \beta}
{\log (m_{{\tilde b}_2}^2 / m_{{\tilde b}_1}^2) \over (m_{{\tilde b}_2}^2 - m_{{\tilde b}_1}^2)} \cr
    M_{13}^b & = &\mbox{} - {3 m_b^4 \lambda A_b \Delta_{{\tilde b}_1} \Delta_{{\tilde b}_2}
\over 8 \pi^2 v \cos \beta \cot \beta}
{g(m_{{\tilde b}_1}^2, \ m_{{\tilde b}_2}^2) \over (m_{{\tilde b}_2}^2 - m_{{\tilde b}_1}^2)^2}
- {3 m_b^4 \lambda \sin \beta \Delta_{{\tilde b}_2} \over 8 \pi^2 v \cos^2 \beta}
{\log (m_{{\tilde b}_2}^2 / m_{{\tilde b}_1}^2) \over (m_{{\tilde b}_2}^2 - m_{{\tilde b}_1}^2) }  , \cr
    M_{14}^b & = & \mbox{} {3 m_b^4 \lambda A_b^2 s \Delta_{{\tilde b}_1} \sin \phi_b
\over 8 \pi^2 v^2 \cos^3 \beta \cos \alpha}
{g(m_{{\tilde b}_1}^2, \ m_{{\tilde b}_2}^2) \over (m_{{\tilde b}_2}^2 - m_{{\tilde b}_1}^2)^2 }
+ {3 m_b^4 \lambda A_b s \sin \phi_b \over 8 \pi^2 v^2 \cos^3 \beta \cos \alpha}
{\log (m_{{\tilde b}_2}^2 / m_{{\tilde b}_1}^2) \over (m_{{\tilde b}_2}^2 - m_{{\tilde b}_1}^2)}  \ , \cr
    M_{23}^b & = & {3 m_b^4 \lambda^2 s \Delta_{{\tilde b}_2}^2 \over 8 \pi^2 v \cos \beta \cot \beta}
{g(m_{{\tilde b}_1}^2, \ m_{{\tilde b}_2}^2) \over (m_{{\tilde b}_2}^2 - m_{{\tilde b}_1}^2)^2 }
- {3 m_b^2 \lambda^2 s \tan \beta \over 8 \pi^2 v \cos \beta}
f(m_{{\tilde b}_1}^2, \ m_{{\tilde b}_2}^2) , \cr
    M_{24}^b & = & \mbox{} - {3 m_b^4 \lambda^2 A_b s^2 \Delta_{{\tilde b}_2} \sin \phi_b
\over 8 \pi^2 v^2 \cos^3 \beta \cos \alpha}
{g(m_{{\tilde b}_1}^2, \ m_{{\tilde b}_2}^2) \over (m_{{\tilde b}_2}^2 - m_{{\tilde b}_1}^2)^2 } \ , \cr
    M_{34}^b & = & \mbox{} - {3 m_b^4 \lambda^2 A_b s \Delta_{{\tilde b}_2} \sin \phi_b
\over 8 \pi^2 v \cos^2 \beta \cot \beta \cos \alpha}
{g(m_{{\tilde b}_1}^2, \ m_{{\tilde b}_2}^2) \over (m_{{\tilde b}_2}^2 - m_{{\tilde b}_1}^2)^2 } \ ,
\end{eqnarray}
where
\begin{eqnarray}
 \Delta_{{\tilde b}_1} & = & A_b - \lambda s \tan \beta \cos \phi_b \ , \cr
 \Delta_{{\tilde b}_2} & = & A_b \cos \phi_b - \lambda s \tan \beta  \  ,
\end{eqnarray}
and the matrix elements of $M^k$ as
\begin{eqnarray}
    M_{11}^k & = & {3 m_k^4 \lambda^2 v^2 \sin^2 \beta \Delta_{{\tilde k}_1}^2 \over 8 \pi^2  s^2 }
{g(m_{{\tilde k}_1}^2, \ m_{{\tilde k}_2}^2) \over (m_{{\tilde k}_2}^2 - m_{{\tilde k}_1}^2)^2}  \ , \cr
    M_{22}^k & = & {3 m_k^4 \lambda^2 v^2 \cos^2 \beta \Delta_{{\tilde k}_1}^2 \over 8 \pi^2 s^2}
{g(m_{{\tilde k}_1}^2, \ m_{{\tilde k}_2}^2) \over (m_{{\tilde k}_2}^2 - m_{{\tilde k}_1}^2 )^2} \ , \cr
    M_{33}^k & = & {3 m_k^4 A_k^2 \Delta_{{\tilde k}_2}^2 \over 8 \pi^2 s^2}
{g(m_{{\tilde k}_1}^2, \ m_{{\tilde k}_2}^2) \over (m_{{\tilde k}_2}^2 - m_{{\tilde k}_1}^2)^2}
 + {3 m_k^4 A_k \Delta_{{\tilde k}_2} \over 4 \pi^2 s^2}
{\log (m_{{\tilde k}_2}^2 / m_{{\tilde k}_1}^2) \over (m_{{\tilde k}_2}^2 - m_{{\tilde k}_1}^2)} \cr
& &\mbox{} + {3 m_k^4 \over 8 \pi^2 s^2}
\log \left ( {m_{{\tilde k}_1}^2  m_{{\tilde k}_2}^2 \over m_k^4} \right ) \ , \cr
    M_{44}^k & = & {3 m_k^4 \lambda^2 A_k^2 v^2 \sin^2 \phi_k \over 8 \pi^2 s^2 \cos^2 \alpha}
{g(m_{{\tilde k}_1}^2, \ m_{{\tilde k}_2}^2) \over (m_{{\tilde k}_2}^2 - m_{{\tilde k}_1}^2 )^2}  \ , \cr
    M_{12}^k & = & {3 m_k^4 \lambda^2 v^2 \sin 2 \beta \Delta_{{\tilde k}_1}^2 \over 16 \pi^2 s^2}
{g(m_{{\tilde k}_1}^2, \ m_{{\tilde k}_2}^2) \over (m_{{\tilde k}_2}^2 - m_{{\tilde k}_1}^2)^2 }
- {3 m_k^2 \lambda^2 v^2 \sin 2 \beta \over 16 \pi^2 s^2}
f(m_{{\tilde k}_1}^2, \ m_{{\tilde k}_2}^2) , \cr
    M_{13}^k & = &\mbox{} - {3 m_k^4 \lambda A_k v \sin \beta \Delta_{{\tilde k}_1} \Delta_{{\tilde k}_2}
\over 8 \pi^2 s^2}
{g(m_{{\tilde k}_1}^2, \ m_{{\tilde k}_2}^2) \over (m_{{\tilde k}_2}^2 - m_{{\tilde k}_1}^2)^2} \cr
& &\mbox{}- {3 m_k^4 \lambda v \sin \beta \Delta_{{\tilde k}_1} \over 8 \pi^2 s^2}
{\log (m_{{\tilde k}_2}^2 / m_{{\tilde k}_1}^2) \over (m_{{\tilde k}_2}^2 - m_{{\tilde k}_1}^2)} \cr
    M_{14}^k & = & \mbox{} - {3 m_k^4 \lambda^2 A_k v^2 \sin \beta \Delta_{{\tilde k}_1} \sin \phi_k
\over 8 \pi^2 s^2 \cos \alpha}
{g(m_{{\tilde k}_1}^2, \ m_{{\tilde k}_2}^2) \over (m_{{\tilde k}_2}^2 - m_{{\tilde k}_1}^2)^2 } \ , \cr
    M_{23}^k & = &\mbox{} - {3 m_k^4 \lambda A_k v \cos \beta \Delta_{{\tilde k}_1} \Delta_{{\tilde k}_2}
\over 8 \pi^2 s^2}
{g(m_{{\tilde k}_1}^2, \ m_{{\tilde k}_2}^2) \over (m_{{\tilde k}_2}^2 - m_{{\tilde k}_1}^2)^2} \cr
& &\mbox{}- {3 m_k^4 \lambda v \cos \beta \Delta_{{\tilde k}_1} \over 8 \pi^2 s^2}
{\log (m_{{\tilde k}_2}^2 / m_{{\tilde k}_1}^2) \over (m_{{\tilde k}_2}^2 - m_{{\tilde k}_1}^2) }  , \cr
    M_{24}^k & = & \mbox{} - {3 m_k^4 \lambda^2 A_k v^2 \cos \beta \Delta_{{\tilde k}_1} \sin \phi_k
\over 8 \pi^2 s^2 \cos \alpha}
{g(m_{{\tilde k}_1}^2, \ m_{{\tilde k}_2}^2) \over (m_{{\tilde k}_2}^2 - m_{{\tilde k}_1}^2)^2 } \ , \cr
    M_{34}^k & = & \mbox{} {3 m_k^4 \lambda A_k^2 v \Delta_{{\tilde k}_2} \sin \phi_k
\over 8 \pi^2 s^2 \cos \alpha}
{g(m_{{\tilde k}_1}^2, \ m_{{\tilde k}_2}^2) \over (m_{{\tilde k}_2}^2 - m_{{\tilde k}_1}^2)^2 }
+ {3 m_k^4 \lambda A_k v \sin \phi_k \over 8 \pi^2 s^2 \cos \alpha}
{\log (m_{{\tilde k}_2}^2 / m_{{\tilde k}_1}^2) \over (m_{{\tilde k}_2}^2 - m_{{\tilde k}_1}^2)}  \ ,
\end{eqnarray}
where
\begin{eqnarray}
 \Delta_{{\tilde k}_1} & = & A_k \cos \phi_k - \lambda v \tan \alpha \  , \cr
 \Delta_{{\tilde k}_2} & = & A_k - \lambda v \tan \alpha \cos \phi_k \ .
\end{eqnarray}

Notice that the matrix elements $M_{i4}^t$, $M_{i4}^b$, and $M_{i4}^k$ ($i=1-3$)
are proportional to $\sin \phi_t$, $\sin \phi_b$, and $\sin \phi_k$, respectively.
If the three CP phases are not all zero, these matrix elements would not be all zero,
and therefore the matrix elements $M_{i4}$ ($i=1-3$) would not be zero,
which are responsible for the scalar-psuedoscalar mixing and hence CP violation.
In short, these three CP phases in the radiative corrections generate
the CP mixing between the scalar and pseudoscalar components,
implying that the four neutral Higgs bosons are not states of definite CP parity.

The four physical neutral Higgs bosons at the one-loop level are defined
as the eigenstates of the squared mass matrix $M$, with their squared masses as the eigenvalues of $M$.
Let us denote the physical four neutral Higgs bosons as $h_i$ ($i$= 1 to 4)
and their squared masses as $m_{h_i}^2$ ($i$ = 1 to 4).
We assume that $m_{h_i}^2 \le m_{h_j}^2$ for $i < j$.
Among them, the upper bound on $m_{h_1}^2$, the squared mass of the lightest neutral Higgs boson,
may be obtained by noticing that the smallest eigenvalue of a positive symmetric matrix
cannot exceed the smaller eigenvalue of its upper left $2\times2$ submatrix.
Thus, the upper bound on $m_{h_1}^2$ at the one-loop level is given as
\begin{eqnarray}
m_{h_1}^2 & \le & \lambda^2 v^2 \sin^2 2 \beta + m_Z^2 \cos^2 2 \beta
+ 2 {g'}^2 v^2 ({\tilde Q}_1 \cos^2 \beta + {\tilde Q}_2 \sin^2 \beta)^2 \cr
& &\mbox{} + {3 m_t^4 \over 8 \pi^2 v^2}
{(\lambda s \cot \beta \Delta_{{\tilde t}_1} - A_t \Delta_{{\tilde t}_2})^2
\over (m_{{\tilde t}_2}^2 - m_{{\tilde t}_1}^2)^2}
g(m_{{\tilde t}_1}^2, \ m_{{\tilde t}_2}^2) \cr
&  & \mbox{} + {3 m_t^4 \over 4 \pi^2 v^2}
{(\lambda s \cot \beta \Delta_{{\tilde t}_1} - A_t \Delta_{{\tilde t}_2})
\over (m_{{\tilde t}_2}^2 - m_{{\tilde t}_1}^2)}
\log \left({m_{{\tilde t}_2}^2 \over m_{{\tilde t}_1}^2}\right)
+ {3 m_t^4 \over 8 \pi^2 v^2}
\log({m_{{\tilde t}_1}^2 m_{{\tilde t}_2}^2 \over m_t^4}) \cr
& &\mbox{} + {3 m_b^4 \over 8 \pi^2 v^2}
{(\lambda s \tan \beta \Delta_{{\tilde b}_1} - A_b \Delta_{{\tilde b}_2})^2
\over (m_{{\tilde b}_2}^2 - m_{{\tilde b}_1}^2)^2}
g(m_{{\tilde b}_1}^2, \ m_{{\tilde b}_2}^2) \cr
&  & \mbox{} + {3 m_b^4 \over 4 \pi^2 v^2}
{(\lambda s \tan \beta \Delta_{{\tilde b}_1} - A_t \Delta_{{\tilde b}_2})
\over (m_{{\tilde b}_2}^2 - m_{{\tilde b}_1}^2)}
\log \left({m_{{\tilde b}_2}^2 \over m_{{\tilde b}_1}^2}\right)
+ {3 m_b^4 \over 8 \pi^2 v^2}
\log({m_{{\tilde b}_1}^2 m_{{\tilde b}_2}^2 \over m_b^4}) \cr
& &\mbox{} + {3 m_k^4 \lambda^2 v^2 \Delta_{{\tilde k}_1}^2 \over 8 \pi^2 s^2}
{g(m_{{\tilde k}_1}^2, \ m_{{\tilde k}_2}^2)
\over (m_{{\tilde k}_2}^2 - m_{{\tilde k}_1}^2)^2}
- {3 m_k^2 \lambda^2 v^2 \sin^2 2 \beta \over 16 \pi^2 s^2} f(m_{{\tilde k}_1}^2, \ m_{{\tilde k}_2}^2) \ ,
\end{eqnarray}
where the first three terms come from the tree-level Higgs potential while the other terms come
from the one-loop corrections due to top quark, bottom quark, the exotic quark and their superparticles.

\section{NUMERICAL ANALYSIS}

For numerical analysis, we assume that the extra $U(1)'$ emerges from the $E_6$ group,
as described in the previous section.
To be concrete, we take the ${\nu}$-model [19], where $U(1)'$ is a mixture of $U(1)_{\chi}$
and $U(1)_{\psi}$ with the mixing angle $\theta_E = \tan^{-1} \sqrt{15}$.
At the electroweak scale, the analysis of renormalization group equation leads to ${g'}_1(m_Z) \simeq 0.46$
for $U(1)'$ gauge coupling constant [19].
Then, the $U(1)'$ charges of $H_1$, $H_2$, and $S$ are given respectively as
${\tilde Q}_1 \approx -0.4910123$, ${\tilde Q}_2 \approx -0.2995571$,
and ${\tilde Q}_3 = - ({\tilde Q}_1 + {\tilde Q}_2)$.
We set quark masses as $m_t$ = 175 GeV, $m_b$ = 4 GeV, and $m_k$ = 700 GeV.
The renormalization scale is set as $\Lambda$ = 700 GeV in the one-loop effective potential.
We assume that the lighter squarks are larger than top quark mass.
For the remaining parameters, we set the ranges for their variations as follows:
$100 \leq A_t = A_b = A_k, {\bar m}_A {\rm ~(GeV)} \leq 2000$,
$100 \leq m_{\rm SUSY} = m_T = m_B = m_K {\rm ~(GeV)}\leq 1000$,
$1500 \leq s {\rm ~(GeV)} \leq 2000$,
$0 \le \phi_t, \phi_b, \phi_k \le 2 \pi$,
$1 < \tan \beta \leq 30$,
and $0 < \lambda \le 0.85$.
Note that one need not vary $\phi_0$ since it appears always together with $A_{\lambda}$
in the definition of ${\bar m}_A$.
Also note that the upper bound on $\lambda$ is chosen by considering
that the the maximum value of $\lambda$ is about 0.83 in the analysis of the renormalization group equation.
A general comment on $s$ may be that a large $s$ leads to a heavy $Z'$ in the present model.
Moreover, the large $s$ can satisfy the experimental constraint
that the mixing angle between $Z$ and $Z'$ should be smaller than 2-3 $\times 10^{-3}$.

We calculate the upper bound on $m_{h_1}$ by varying all the relevant parameters
within their allowed ranges for a given $\tan\beta$.
The results are shown in Fig. 1, where the upper bound on $m_{h_1}$ is plotted
as a function of $\tan \beta$,
The solid curve is obtained by considering only the tree-level Higgs potential,
whereas the dashed curve is obtained by considering the full one-loop Higgs potential
with explicit CP violation.
It is observed in the MSSM that the lightest neutral Higgs boson mass increases steadily
as $\tan \beta$ increases.
In the present model, where the MSSM is incorporated with an extra $U(1)'$,
the lightest neutral Higgs boson mass does not increase in accordance with the increase of $\tan \beta$,
because of both the Higgs singlet and $V_D$ contributions due to an extra $U(1)'$.
One can see that in Fig. 1 the lightest neutral Higgs boson mass becomes maximum
at $\tan \beta \sim 1.5$ for both curves.

As we have remarked in the previous section, the scalar-pseudoscalar mixing,
which is responsible for CP violation in this model, is triggered by three CP phases,
$\phi_t$, $\phi_b$, and $\phi_k$.
The size of the CP mixing between the scalar and pseudoscalar Higgs bosons could be regarded as maximal
if $\sin \phi_t =\sin \phi_b =\sin \phi_k$ = 1, while there would be no CP mixing
if $\sin \phi_t =\sin \phi_b =\sin \phi_k$ = 0.
In order to see a more clear picture about the amount of the CP mixing among the scalar
and pseudoscalar Higgs bosons, we introduce ($i$ = 1,2,3)
\begin{equation}
    \omega_i (\phi_k)
    = {2 |M_{i4}| \over |M_{ii}| + |M_{44}|} \ ,
\end{equation}
which depend implicitly on other parameters as well.

These $\omega_i (\phi_k)$ measure by what amount $\phi_k$ contributes to the size of CP mixing.
In order to express the relative contributions by $\phi_k$, we further introduce
\begin{equation}
    \Omega_i (\phi_k)
    = {\omega_i(\phi_k) \over \omega_i(\phi_k=0)} \ .
\end{equation}
By evaluating  $\Omega_i (\phi_k)$, we may study the role of the exotic quark sector
in the CP violation in the present model.

In Fig. 2, we plot $\Omega_3$ as a function of $\phi_k$,
where we fix $m_{\rm SUSY} = {\bar m}_A = A_t = s/2= 1000$ GeV and $\lambda = 0.3$,
and we set for simplicity $\phi_t = \phi_b$.
The four curves in Fig. 2 correspond to four different sets of $\phi_t = \phi_b$ and $\tan \beta$,
namely, the solid curve is obtained for $\phi_t = \phi_b = \pi/6$ and $\tan \beta = 2$,
the dashed curve for $\phi_t = \phi_b = \pi/6$ and $\tan \beta = 20$,
the dotted curve for $\phi_t = \phi_b = \pi/3$ and $\tan \beta = 2$, and
the dash-dotted curve for $\phi_t = \phi_b = \pi/3$ and $\tan \beta = 20$.
It is trivial to notice that $\Omega_3 = 1$ at $\phi_k = 0$, $\pi$, or $2\pi$.
For other values of $\phi_k$, we find that $\Omega_3$ may change
up to as much as about $5 \times 10^3$ for some parameter values.
This implies that the effect of $\phi_k$ upon $\Omega_3$, or $\omega_3$, is very significant.
In other words, the matrix element $|M_{34}|$ depends critically
on the complex phase of the exotic quark sector.

We study the behaviors of $\Omega_1$ and $\Omega_2$ for $0 \le \phi_k \le 2\pi$,
fixing the values of other parameters as in Fig. 2.
Unlike $\Omega_3$, they do not fluctuate widely against the variation of $\phi_k$.
Thus, it can be said that the complex phase of the exotic quark sector has
a strong influence on the magnitude of the mixing between $h_3$ and $h_4$,
whereas it has relatively weak influence on the magnitudes of
the mixing between $h_1$ and $h_4$ or between $h_2$ and $h_4$.

We calculate the masses of the neutral Higgs bosons for the same parameter setting as in Fig. 2:
$m_{\rm SUSY} = {\bar m}_A = A_t = s/2= 1000$ GeV, $\lambda = 0.3$,
together with four different sets of  $\phi_t = \phi_b$ and $\tan \beta$, for $0 \le \phi_k \le 2\pi$.
We obtain approximately 126, 1000, 1002, and 1043 GeV, respectively,
for $m_{h_1}$, $m_{h_2}$, $m_{h_3}$, and $m_{h_4}$.
We find that these values are almost unchanged for $0 \le \phi_k \le 2\pi$,
and nearly the same within a few GeV  for the four different sets of  $\phi_t = \phi_b$ and $\tan \beta$.

The mass of the extra gauge boson $Z'$ is also estimated.
We find that $m_{Z'}$ is stable at approximately 1039 GeV for the parameter values in Fig. 2.
The mixing angle between $Z$ and $Z'$, denoted as $\alpha_{ZZ'}$, depends on both  $\tan \beta$ and $s$.
We find that $\alpha_{ZZ'} = 1.39 \times 10^{-3}$ for $\tan \beta =2$,
and and slightly increases for larger values of $\tan\beta$ as $2.93 \times 10^{-3}$ for $\tan \beta =20$.

We have studied the dependence of the mixing element between $h_3$ and $h_4$
on the complex phase of the exotic quark sector.
Meanwhile, one can evaluate effectively the absolute size of the CP violation
by calculating the dimensionless parameter
\begin{equation}
    \rho (\phi_t, \phi_b, \phi_k)  = 4 \sqrt{|O_{11} O_{21} O_{31} O_{41}|} \ ,
\end{equation}
where $O_{ij}$ are the elements of the orthogonal matrix which diagonalizes the squared mass matrix
for the neutral Higgs bosons.
The range of $\rho$ goes from 0 to 1 since the elements of the orthogonal matrix satisfy
the orthogonality condition of $\sum_{j = 1}^4 O_{j1}^2 = 1$.
If $\rho = 0$, there would be no CP violation, whereas the CP symmetry would be maximally violated if $\rho = 1$.
The maximal CP violation that leads to $\rho$ = 1 takes place when $O_{11}^2 = O_{21}^2 = O_{31}^2 = O_{41}^2 = 1/4$.

In order to figure out the dependence of $\rho$ on $\phi_k$, we introduce
\begin{equation}
    \rho_k (\phi_k) = {\rho (\phi_t, \phi_b, \phi_k)  \over \rho(\phi_t, \phi_b, \phi_k = 0)}
\end{equation}
We calculated $\rho_k$ as a function of $\phi_k$, for the same parameter setting as Fig. 2.
If $\rho_k$ remains at 1, it would imply that $\rho_k$ does not depend on $\phi_k$.
On the other hand, $\phi_k$ would contribute more significantly to CP violation
if $\rho_k$ moves farther away from 1.
The result is shown in Fig. 3.
The four curves correspond to the four sets of parameter values.
Fig. 3 shows that the fluctuation of $\rho_k$ is larger for $\tan \beta = 20$ than for $\tan \beta = 2$.
This implies that $\phi_k$ play an important role in $\rho_k$ for large $\tan \beta$.
One can see that $\rho_k$ fluctuates by up to 20 \% for nonzero $\phi_k$.
Since $\rho(\phi_t, \phi_b, \phi_k)$ is an absolute measure for the CP violation
in our model, it is reasonable to expect that the complex phase of the exotic quark sector may
change up to 20 \% of the CP violation.

\section{SPONTANEOUS CP VIOLATION}

In this section, let us examine briefly whether it is possible
to realize the spontaneous CP violation in the present model.
At initial stage, the Lagrangian density of the model is assumed to be invariant
with respect to CP property.
After the electroweak symmetry breaking, the CP symmetry may spontaneously be broken
by complex phases in the VEVs of the neutral Higgs fields.
In the MSSM with an extra $U(1)'$, one can conjecture from the Georgi-Pais theorem
that the spontaneous CP violation leads to a very light pseudoscalar Higgs boson, namely, the axion.
It has been addressed that the spontaneous CP violation through radiative corrections can be realized
when at the tree level there exist massless Higgs bosons other than Goldstone boson [23].
In the present model, there is no pseudo-Glodstone boson, which is referred to as axion,
and thus the spontaneous CP violation cannot occur at the one-loop level.

In order to examine the possibility of spontaneous CP violation scenario at the one-loop level,
we can quantitatively analyze the axion mass by using the formulae which are derived
for explicit CP violation scenario.
In spontaneous CP violation scenario, the one-loop effective potential of the present model
may have only one physical CP phase ($\phi_s$) arising from the VEV of the Higgs singlet
by considering the radiative corrections due to
top quark, bottom quark, the exotic quark, and their superpartners.
At the one-loop level, we obtain the minimum condition for the vacuum with respect to $\phi_s$ as
\begin{eqnarray}
A_{\lambda} & = & {3 m_t^2 A_t \over 16 \pi^2 v^2 \sin^2 \beta} f (m_{{\tilde t}_1}^2,  \ m_{{\tilde t}_2}^2)
+ {3 m_b^2 A_b \over 16 \pi^2 v^2 \cos^2 \beta} f (m_{{\tilde b}_1}^2,  \ m_{{\tilde b}_2}^2) \cr
&&\mbox{}+  {3 m_k^2 A_k \over 16 \pi^2 s^2} f (m_{{\tilde k}_1}^2,  \ m_{{\tilde k}_2}^2)
\end{eqnarray}
for $\phi_s \neq 0, \pi$ .
We obtain the axion mass at the one-loop level as
\begin{eqnarray}
m_a^2 & =& {3 m_t^4 \lambda^2 A_t^2 s^2 \sin^2 \phi_t \over 8 \pi^2 v^2 \sin^4 \beta \cos^2 \alpha}
{g(m_{{\tilde t}_1}^2, \ m_{{\tilde t}_2}^2) \over (m_{{\tilde t}_2}^2 - m_{{\tilde t}_1}^2 )^2}
+ {3 m_b^4 \lambda^2 A_b^2 s^2 \sin^2 \phi_b \over 8 \pi^2 v^2 \cos^4 \beta \cos^2 \alpha}
{g(m_{{\tilde b}_1}^2, \ m_{{\tilde b}_2}^2) \over (m_{{\tilde b}_2}^2 - m_{{\tilde b}_1}^2 )^2}  \cr
&  &\mbox{} + {3 m_k^4 \lambda^2 A_k^2 v^2 \sin^2 \phi_k \over 8 \pi^2 s^2 \cos^2 \alpha}
{g(m_{{\tilde k}_1}^2, \ m_{{\tilde k}_2}^2) \over (m_{{\tilde k}_2}^2 - m_{{\tilde k}_1}^2 )^2}  \ .
\end{eqnarray}
The massless function $g$ in the above expression is found to have negative or very small values
for the entire parameter space we consider, in the case of explicit CP violation scenario.
Employing this result from the explicit CP violation scenario, the squared mass of axion
is almost always negative for the entire parameter space, which is unacceptable.
Therefore, the spontaneous CP violation is not viable for the MSSM with an extra $U(1)'$.

Furthermore, the lightest neutral Higgs boson mass also faces difficulty
in the spontaneous CP violation scenario.
Calculate the lightest neutral Higgs boson  mass from the $4 \times 4$ squared mass matrix
for neutral Higgs bosons, we find that the squared mass of the lightest Higgs boson has negative values
for the above parameter space.
In fact, the minimum condition for the vacuum with respect to $\phi_s$ has a very strong constraint,
and thus the vacuum becomes unstable for the full parameter space we consider,
in the spontaneous CP violation scenario.

\section{CONCLUSIONS}

We consider CP violation in the MSSM with an extra $U(1)'$ symmetry, which is
assumed to be originated from a string-inspired $E_6$ model.
The charge of the extra $U(1)'$ is defined as $Q' = \cos \theta_E Q_{\chi} + \sin \theta_E Q_{\psi}$
with $\theta_E = \tan^{-1}\sqrt{15}$.
This model possesses one Higgs singlet and exotic quarks and squarks,
in addition to the particle content of the MSSM.
In this model, the CP violation is impossible to occur either explicitly or spontaneously
in its the tree-level Higgs sector.
The mixing between the scalar and pseudoscalar Higgs bosons, which induce the CP violation,
may occur via the radiative corrections at the one-loop level.
For the radiative corrections, we consider the loop contributions due to top quark,
bottom quark, the exotic quark and their superpartners.
The radiative corrections contain in general three complex phases originated
from the top squark sector, the bottom squark sector, and the exotic squark sector.
We pay attention to the complex phase $\phi_k$ that comes from the exotic squark sector.

We calculate the lightest neutral Higgs boson mass.
In explicit CP violation scenario, we find that the mass of the lightest neutral Higgs boson
is smaller than about 135 GeV and 165 GeV, respectively, at the tree level and at the one-loop level,
for the whole parameter space that we considered.

For four different sets of parameters, we calculate the magnitude of
the CP mixing among the scalar and pseudoscalar Higgs bosons for $0 \le \phi_k \le 2\pi$.
The matrix element, $|M_{34}|$,
representing the CP mixing between the heaviest scalar and the pseudoscalar Higgs bosons,
is found to depend heavily on the complex phase from the exotic squark sector, $\phi_k$.
In particular, the magnitude of the CP mixing between the heaviest scalar and the pseudoscalar Higgs bosons
is found to change as much as about 20 \% for $ 0 \le \phi_k \le 2 \pi$
when $\phi_t = \phi_b = \pi/3$ and $\tan \beta = 20$, whereas the CP mixing among
the other scalar Higgs bosons and the pseudoscalar Higgs boson
is found to be relatively stable against the variation of $\phi_k$.

Also, we have investigated the possibility of spontaneous CP violation at the one-loop level
in the neutral Higgs sector, where the tree-level Higgs potential possesses CP symmetry.
We argue that it is impossible to achieve spontaneous CP violation in the radiatively corrected Higgs potential
of this model.

\vskip 0.3 in
\noindent
{\large {\bf ACKNOWLEDGMENTS}}
\vskip 0.2 in
This research is supported by KOSEF through CHEP.
The authors would like to acknowledge the support from KISTI
(Korea Institute of Science and Technology Information) under
"The Strategic Supercomputing Support Program" with Dr. Kihyeon Cho as the technical supporter.
The use of the computing system of the Supercomputing Center is also greatly appreciated.



\vfil\eject

{\large {\bf FIGURE CAPTION}}
\vskip 0.3 in
\noindent
FIG. 1. : The upper bound on the lightest neutral Higgs boson mass as a function of $\tan \beta$,
where other relevant parameters are allowed to vary within their respective ranges:
$100 \leq A_t = A_b = A_k, {\bar m}_A {\rm ~(GeV)} \leq 2000$,
$100 \leq m_{\rm SUSY} = m_T = m_B = m_K {\rm ~(GeV)}\leq 1000$,
$1500 \leq s {\rm ~(GeV)} \leq 2000$,
$0 \le \phi_t, \phi_b, \phi_k \le 2 \pi$,
$1 < \tan \beta \leq 30$,
and $0 < \lambda \le 0.85$.
The quark masses are set as $m_t$ = 175 GeV, $m_b$ = 4 GeV, $m_k$ = 700 GeV,
and the renormalization scale is set as $\Lambda$ = 700 GeV.
The solid and dashed curves are respectively obtained from the tree-level
and the one-loop level Higgs potential with explicit CP violation.

\vskip 0.3 in
\noindent
FIG. 2. : Plot of $\Omega_3$ as a function of $\phi_k$,
for four different sets of $\phi_t = \phi_b$ and $\tan \beta$:
$\phi_t = \phi_b = \pi/6$ and $\tan \beta = 2$ (solid curve),
$\phi_t = \phi_b = \pi/6$ and $\tan \beta = 20$ (dashed curve),
$\phi_t = \phi_b = \pi/3$ and $\tan \beta = 2$ (dotted curve), and
$\phi_t = \phi_b = \pi/3$ and $\tan \beta = 20$ (dash-dotted curve).
The remaining parameters are set as $m_{\rm SUSY} = {\bar m}_A = A_t = s/2= 1000$ GeV and $\lambda = 0.3$.

\vskip 0.3 in
\noindent
FIG. 3. : Plot of $\rho_k$ as a function of $\phi_k$, for the same parameter setting as Fig. 2.

\vfil\eject

\setcounter{figure}{0}
\def\figurename{}{}%
\renewcommand\thefigure{FIG. 1}
\begin{figure}[t]
\begin{center}
\includegraphics[scale=0.6]{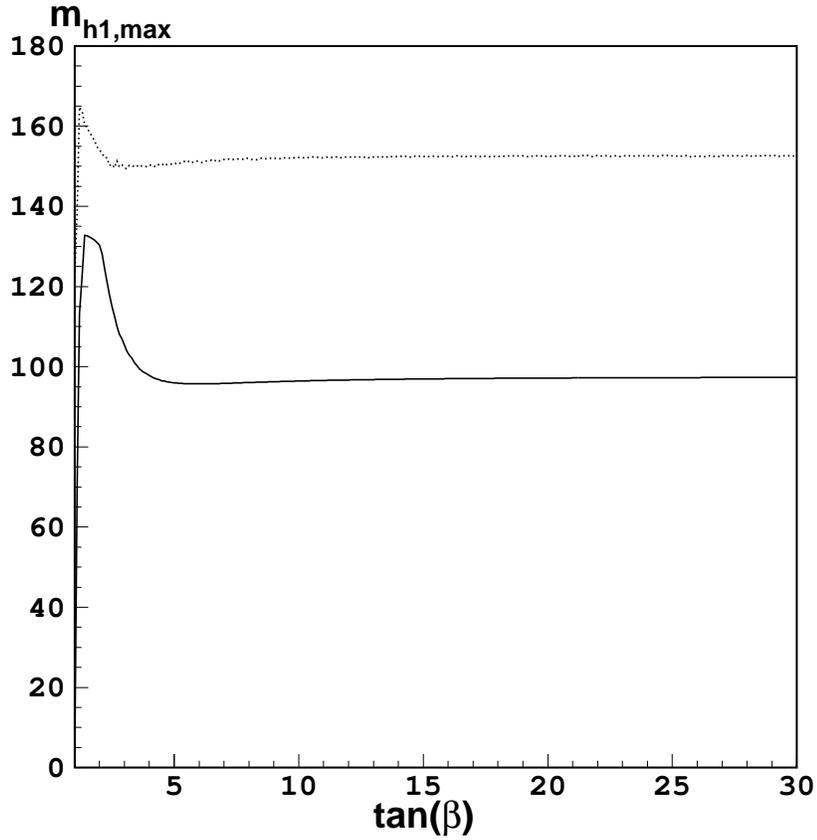}
\caption[plot]{The upper bound on the lightest neutral Higgs boson mass as a function of $\tan \beta$,
where other relevant parameters are allowed to vary within their respective ranges:
$100 \leq A_t = A_b = A_k, {\bar m}_A {\rm ~(GeV)} \leq 2000$,
$100 \leq m_{\rm SUSY} = m_T = m_B = m_K {\rm ~(GeV)}\leq 1000$,
$1500 \leq s {\rm ~(GeV)} \leq 2000$,
$0 \le \phi_t, \phi_b, \phi_k \le 2 \pi$,
$1 < \tan \beta \leq 30$,
and $0 < \lambda \le 0.85$.
The quark masses are set as $m_t$ = 175 GeV, $m_b$ = 4 GeV, $m_k$ = 700 GeV,
and the renormalization scale is set as $\Lambda$ = 700 GeV.
The solid and dashed curves are respectively obtained from the tree-level
and the one-loop level Higgs potential with explicit CP violation.}
\end{center}
\end{figure}

\renewcommand\thefigure{FIG. 2}
\begin{figure}[t]
\begin{center}
\includegraphics[scale=0.6]{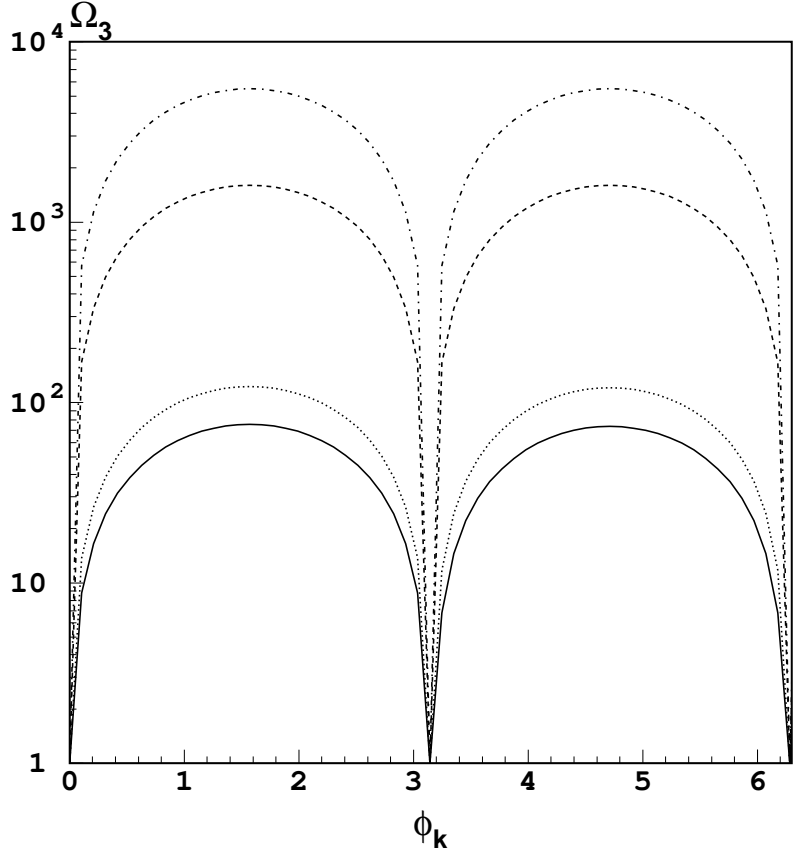}
\caption[plot]{Plot of $\Omega_3$ as a function of $\phi_k$,
for four different sets of $\phi_t = \phi_b$ and $\tan \beta$:
$\phi_t = \phi_b = \pi/6$ and $\tan \beta = 2$ (solid curve),
$\phi_t = \phi_b = \pi/6$ and $\tan \beta = 20$ (dashed curve),
$\phi_t = \phi_b = \pi/3$ and $\tan \beta = 2$ (dotted curve), and
$\phi_t = \phi_b = \pi/3$ and $\tan \beta = 20$ (dash-dotted curve).
The remaining parameters are set as $m_{\rm SUSY} = {\bar m}_A = A_t = s/2= 1000$ GeV and $\lambda = 0.3$.}
\end{center}
\end{figure}

\renewcommand\thefigure{FIG. 3}
\begin{figure}[t]
\begin{center}
\includegraphics[scale=0.6]{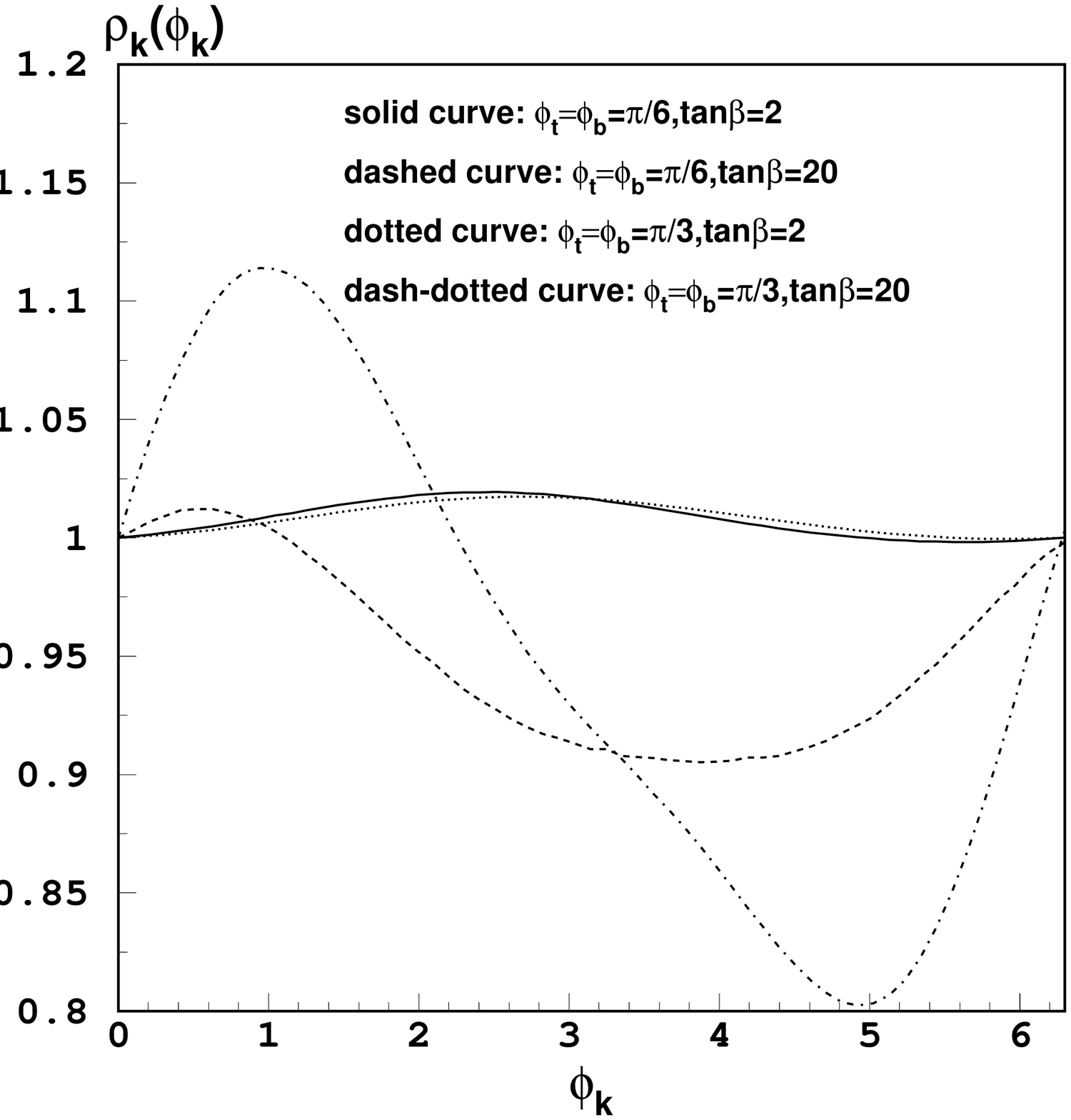}
\caption[plot]{Plot of $\rho_k$ as a function of $\phi_k$, for the same parameter setting as Fig. 2.}
\end{center}
\end{figure}


\begin{thebibliography}{99}
\bibitem{1} H.P. Nilles, Phys. Rep. {\bf 110}, 1 (1984);
    J.F. Gunion, H.E. Haber, G.L. Kane, and S. Dawson,
    {\it The Higgs Hunters' Guide} (Addison-Wesley Redwood City, CA, 1990).
\bibitem{2} S. Weinberg, Phys. Rev. Lett. {\bf 37}, 657 (1976).
\bibitem{3} A. Pomarol, Phys. Lett. B {\bf 287}, 331 (1992);
    N. Maekawa, Phys. Lett. B {\bf 282}, 387 (1992);
    N. Haba, Phys. Lett. B {\bf 398}, 305 (1997).
\bibitem{4} O. Lebedev, Eur. Phys. J. C {\bf 4}, 363 (1998).
\bibitem{5} A. Pilaftsis, Phys. Lett. B {\bf 435}, 88 (1998);
    A. Pilaftsis, Phys. Rev. D {\bf 58}, 096010 (1998).
\bibitem{6} D.A. Demir, Phys. Rev. D {\bf 60}, 055006 (1999);
    S.Y. Choi, M. Drees, and J.S. Lee, Phys. Lett. B {481}, 57 (2000).
\bibitem{7} G.L. Kane and L.T. Wang, Phys. Lett. B {\bf 488}, 383 (2000).
\bibitem{8} S. Heinemeyer, Eur. Phys. J. C {\bf 22}, 521 (2001).
\bibitem{9} T. Ibrahim and P. Nath, Phys. Rev. D {\bf 63}, 035009 (2001);
    T. Ibrahim and P. Nath, Phys. Rev. D {\bf 66}, 015005 (2002);
    S.W. Ham, S.K. Oh, E.J. Yoo, C.M. Kim, and D. Son, Phys. Rev. D {\bf 68}, 055003 (2002).
\bibitem{10} J.S. Lee, A. Pilaftsis, M. Carena, S.Y. Choi, M. Drees, J.R. Ellis, and, C.E.M. Wagner,
    Comput. Phys. Commun. {\bf 156}, 283 (2004).
\bibitem{11} J.R. Ellis, J.S. Lee, and A. Pilaftsis, Phys. Rev. D {\bf 72}, 095006 (2005).
\bibitem{12} A. Pilaftsis and C.E.M. Wagner, Nucl. Phys. {\bf B553}, 3 (1999);
    M. Carena, J. Ellis, A. Pilaftsis, and C.E.M. Wagner, Nucl. Phys. {\bf B586}, 92 (2000);
    S.W. Ham, S.K. Oh, E.J. Yoo, and H.K. Lee, J. Phys. G {\bf 27}, 1 (2001);
    M. Carena, J.R. Ellis, A. Pilaftsis, and C.E.M. Wagner, Phys. Lett. B {\bf 495}, 155 (2000).
\bibitem{13} J.E. Kim and H.P. Nilles, Phys. Lett. B {\bf 138}, 150 (1984).
\bibitem{14} J. Ellis, J.F. Gunion, H.E. Haber, L. Roszkowski, and F. Zwirner, Phys. Rev. D {\bf 39}, 844, (1989).
\bibitem{15} J.L. Hewett and T.G. Rizzo, Phys. Rep. {\bf 183}, 193 (1989);
             A. Leike, Phys. Rep. {\bf 317}, 143 (1999).
\bibitem{16} M. Cvetic and P. Langacker, Phys. Rev. D {\bf 54}, 3570 (1996).
\bibitem{17} M. Cvetic, D.A. Demir, J.R. Espinosa, L. Everett, and P. Langacker,
             Phys. Rev. D {\bf 54}, 3570 (1996).
\bibitem{18} D.A. Demir and N.K. Pak, Phys. Rev. D {\bf 57}, 6609 (1998);
             Y. Daikoku and D. Suematsu, Phys. Rev. D {\bf 62}, 095006 (1998);
             H. Amini, New J. Phys. {\bf 5}, 49 (2003).
\bibitem{19} S.F. King, S. Moretti, and R. Nevzorov, Phys. Rev. D {\bf 73}, 035009 (2006);
             S.F. King, S. Moretti, and R. Nevzorov, Phys. Lett. B {\bf 634}, 278 (2006).
\bibitem{20} D.A. Demir and L.L. Everett, Phys. Rev. D {\bf 69}, 015008 (2004).
\bibitem{21} J. Erler, Nucl. Phys. {\bf B586}, 73 (2000).
\bibitem{22} S. Coleman and E. Weinberg, Phys. Rev. D {\bf 7}, 1888 (1973).
\bibitem{23} H. Georgi and A. Pais, Phys. Rev. D {\bf 10}, 1246 (1974).
\end{thebibliography}
\end{document}